\documentclass[fleqn,12pt]{wlscirep}
\usepackage[utf8]{inputenc}
\usepackage[T1]{fontenc}
\usepackage{graphicx}
\usepackage{amsmath}
\usepackage{amssymb}
\usepackage{booktabs}
\usepackage{lineno}
\usepackage[misc]{ifsym}
\usepackage{hyperref}
\usepackage{multirow}
\graphicspath{{figures/}}
\usepackage{lineno}
\usepackage{diagbox}
\usepackage[normalem]{ulem}
\useunder{\uline}{\ul}{}
\usepackage{float}
\usepackage{caption}
\usepackage{indentfirst}
\usepackage{threeparttable}
\usepackage{wrapfig}
\usepackage{graphicx} 

\title{Large-vocabulary forensic pathological analyses via prototypical cross-modal contrastive learning}

\author[1,+]{Chen Shen}
\author[2,6,+,*]{Chunfeng Lian}
\author[1]{Wanqing Zhang}
\author[3]{Fan Wang}
\author[4]{Jianhua Zhang}
\author[1]{Shuanliang Fan}
\author[1]{Xin Wei}
\author[1]{Gongji Wang}
\author[3]{Kehan Li}
\author[5]{Hongshu Mu}
\author[1]{Hao Wu}
\author[1]{Xinggong Liang}
\author[3,6,*]{Jianhua Ma}
\author[1,*]{Zhenyuan Wang}

\affil[1]{Key Laboratory of National Ministry of Health for Forensic Sciences, School of Medicine \& Forensics, Health Science Center,Xi’an Jiaotong University, Xi'an, Shaanxi 710049, China}
\affil[2]{School of Mathematics and Statistics, Xi'an Jiaotong University, Xi'an, Shaanxi 710149, China}
\affil[3]{Key Laboratory of Biomedical Information Engineering of Ministry of Education, School of Life Science and Technology, Xi'an Jiaotong University, Xi'an, Shaanxi 710049, China}
\affil[4]{Shanghai Key Laboratory of Forensic Medicine, Shanghai Forensic Service Platform, Academy of Forensic Science, Shanghai 200063, China}
\affil[5]{Weicheng Branch, Xian'yang Public Security Bureau, Xian'yang, Shaanxi 710049, China}
\affil[6]{Pazhou Lab (Huangpu), Guangzhou 510000, China}
\affil[+]{These authors contributed equally to this work.} 
\affil[*]{Corresponding authors.}
\begin{abstract}
	Forensic pathology is critical in determining the cause and manner of death through post-mortem examinations, both macroscopic and microscopic.
	The field, however, grapples with issues such as outcome variability, laborious processes, and a scarcity of trained professionals.
	This paper presents SongCi, an innovative visual-language model (VLM) designed specifically for forensic pathology.
	SongCi utilizes advanced prototypical cross-modal self-supervised contrastive learning to enhance the accuracy, efficiency, and generalizability of forensic analyses.
	It was pre-trained and evaluated on a comprehensive multi-center dataset, which includes over $16$ million high-resolution image patches, $2,228$ vision-language pairs of post-mortem whole slide images (WSIs) and corresponding gross key findings, along with $471$ distinct diagnostic outcomes. 
	Our findings indicate that SongCi surpasses existing multi-modal AI models in many forensic pathology tasks, performs comparably to experienced forensic pathologists and significantly better than less experienced ones, and provides detailed multi-modal explainability, offering critical assistance in forensic investigations. 
	To the best of our knowledge, SongCi is the first VLM specifically developed for forensic pathological analysis and the first large-vocabulary computational pathology (CPath) model that directly processes gigapixel WSIs in forensic science.
	\textcolor{blue}{The source code will be released on  \href{https://github.com/shenxiaochenn/SongCi}{https://github.com/shenxiaochenn/SongCi}}.

\end{abstract}
\begin{document}
	
	\flushbottom
	\maketitle
	%
	%
	\thispagestyle{empty}

	\section*{Introduction} \label{sec:intro}
The medicolegal autopsy, commonly known as a post-mortem examination, is conducted by a forensic pathologist to examine the body of a deceased individual meticulously\cite{COVID-19,COVID-autopsy}. The purpose of this examination is to determine the cause and manner of death and identify any diseases or injuries present\cite{autopsy-post}. As a cornerstone of forensic science, autopsies are crucial for both legal and medical analysis\cite{SARS-CoV-2}. Within the criminal justice system, they provide essential evidence that may implicate or exonerate individuals\cite{autopsies_essential}. Additionally, autopsies significantly advance medical knowledge regarding various health conditions\cite{autopsies_contribute}.


Forensic pathologists conduct autopsies using a comprehensive methodology that spans from macroscopic to microscopic analysis. This process encompasses external inspection of the body's surface, internal examination via dissection to assess organ systems, toxicological analysis of bodily fluids, and histopathological analysis of tissue samples\cite{Postmortem examination,COVID-autopsy}. The insights gleaned from these assessments, including macroscopic observations at the organ level and microscopic details at the cellular level, are integral to generating precise and dependable autopsy reports. These reports are essential not only for determining the cause and manner of death but also for estimating the time since death\cite{autopsy-c19}.


However, conducting precise post-mortem examinations, especially in forensic pathology, poses substantial challenges. The accuracy of these examinations relies heavily on the expertise and subjective assessments of forensic pathologists, leading to considerable variability in outcomes\cite{variance_1,variance_2}. Such variability can result in inconsistent findings, even among forensic pathologists with similar training, and is especially marked in complex cases\cite{variance_in,variance_in2}. Forensic pathology is also a labor-intensive and time-consuming discipline, requiring experienced pathologists to invest significant time in analyzing a single whole slide image (WSI)\cite{time}. The complexity increases when multiple organ analyses are necessary. Furthermore, the stringent standards of forensic pathology, combined with a shortage of skilled professionals, exacerbate these issues, impacting the overall efficiency and precision of the field\cite{train_human2,train_human1}.

In recent years, computational pathology (CPath), augmented by artificial intelligence (AI), has demonstrated significant potential in a range of clinical pathology tasks, including cancer diagnosis and subtyping\cite{subtyping,subtyping2}, metastasis detection\cite{metastasis_detection}, and patient survival prediction\cite{mcat}. The typical approach involves training deep neural networks driven by specific tasks on carefully labeled samples. However, the arduous and expensive process of sample collection and labeling, particularly for WSIs, restricts the availability of training data, thereby limiting the scalability and generalizability of these AI models. Inspired by the remarkable advancements in self-supervised learning (SSL) and foundation models within the broader machine learning community, recent CPath studies have begun to pre-train models using a variety of unlabeled data, subsequently fine-tuning them for specific downstream tasks, exemplifying the SSL-based transfer learning paradigm \cite{s_Cross-scale,s_dd}. To enhance generalization and robustness, some innovative studies have integrated pathological images with linguistic information (e.g., pathologists' reports, scholarly articles, or medical textbooks) to pre-train visual-language models (VLMs)\cite{m_Quilt-LLaVA,bottle-shared,m_five}. These models capitalize on the critical semantic information and in-depth domain knowledge contained in textual descriptions to improve the contextual interpretation of histopathological images, resulting in more prosperous and nuanced feature representations. Such advancements in CPath research have significantly refined the workflow and contributed to the advancement of clinical pathology, offering valuable insights for forensic pathological analysis.


Nevertheless, the direct application of clinical-focused models to forensic pathology presents challenges due to the unique characteristics of forensic samples and tasks. Forensic data often exhibit complex post-mortem changes that are absent in clinical biopsies. Moreover, forensic analyses typically encompass a more comprehensive range of conditions (e.g., trauma, disease, and postmortem changes) and organs (e.g., brain, heart, and lung), whereas clinical biopsies usually concentrate on a single organ and task\cite{FP1,FP2}. Consequently, forensic CPath displays more extensive large-vocabulary attributes. These distinctions necessitate the development of specialized models for forensic CPath, which requires forensic-specific pretraining data and more sophisticated SSL approaches to acquire fine-grained, multi-modal representations from such demanding data.

\begin{figure}[h]
		\centering
		\includegraphics[width=\linewidth]{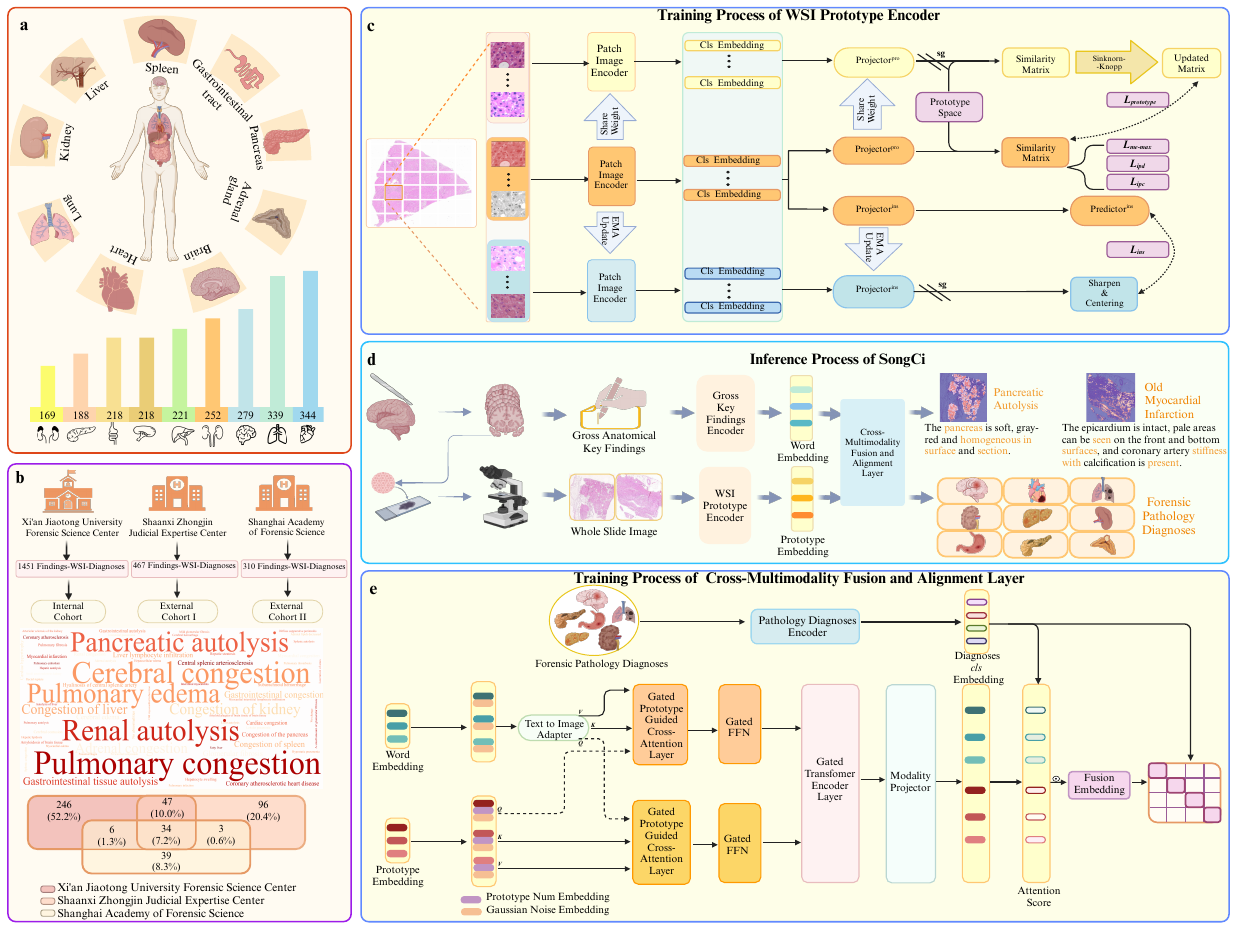}
		\captionsetup{font={tiny,bf,stretch=1.25},justification=raggedright}
		\caption{\textcolor{black}{The framework of SongCi and studied large-vocabulary, multi-center datasets.				
		a. Overview of WSI data. ~The dataset spans a broad spectrum of samples from nine various human organs, each meticulously annotated.
		b. Data Structure and Provenance. ~The dataset was compiled from three premier forensic cohorts. Diagnostic outcomes in forensic pathology were represented using a word cloud and a Venn diagram to illustrate the distribution and overlap of diagnoses.
		c. Training process of WSI prototype encoder. ~SongCi utilizes a self-supervised contrastive learning framework, augmented with prototype-based clustering strategies to enhance efficiency, forming the basis of the prototypical WSI encoder.
		d. Inference process of SongCi. ~SongCi processes the gross anatomical key findings and WSIs to generate a potential diagnosis. Additionally, SongCi provides a diagnostic rationale by highlighting significant terms related to the gross key findings and identifying suspicious regions within the WSIs.
		e. Training process of cross-multimodality fusion and alignment layer. ~Within the SongCi framework, diverse data modalities, including gross key findings and WSIs, are integrated using an innovative gated-attention-boosted multimodal fusion block. Subsequently, the framework aligns the unified representation space with forensic pathological diagnoses through self-supervised contrastive learning, effectively establishing inter-modal correlations.
		For more detailed information, please refer to the Methods section.}}
		\label{fig:songci}
\end{figure}

Addressing these deficiencies, this paper introduces a novel VLM tailored for extensive forensic pathology lexicons. It is pre-trained using cross-modal self-supervised contrastive learning methods on a heterogeneous collection of multi-organ post-mortem WSIs paired with descriptive texts. The model was christened \textbf{SongCi} in tribute to Song Ci, the trailblazer of forensic science during the Southern Song dynasty (Note: Song Ci was a distinguished forensic medical scientist from the Southern Song dynasty, renowned as the inaugural forensic entomologist. His judicial case examinations and experiences are documented in the seminal work ‘Collected Cases of Injustice Rectifie’, available on \href{https://en.wikipedia.org/wiki/Song_Ci}{Wikipedia}.), aiming to augment forensic examinations' precision and efficiency. SongCi strives to deliver more comprehensive and dependable diagnostic results by integrating multi-modal data. Specifically, we collect a multi-center dataset consisting of a total of 2,228 vision-language pairs of post-mortem WSIs (images), gross key findings at the organ level (texts), and final forensic diagnostic outcomes (texts) from nine different organs (Fig.~\ref{fig:songci}a). These data contribute to more than 16 million high-resolution image patches and 471 different diagnostic outcomes, with only 34 of these diagnoses consistently existing across all forensic centers (Fig.~\ref{fig:songci}b). Two pivotal SSL strategies have been developed for nuanced post-mortem representation learning. \textit{First}, we propose a prototypical contrastive learning strategy (Fig.~\ref{fig:songci}c) to construct a prototypical WSI encoder. This encoder transforms all patches of an ultra-high resolution WSI (e.g., $10,000\times10,000$ pixels) into a lower-dimensional prototype space, effectively distilling redundant information to extract generalizable patch representations of post-mortem tissues from various organs. \textit{Second}, leveraging the pre-trained prototypical WSI encoder and a pathology-specific language model~\cite{plip}, we introduce a cross-modal contrastive learning strategy (Fig.~\ref{fig:songci}e). This strategy creates a gated-attention-boosted multi-modal block that integrates representations from paired WSI and gross key findings to align with forensic examination outcomes. Subsequently, the pre-trained prototype encoder, language model, and multi-modal block are integrated for zero-shot inference (Fig.~\ref{fig:songci}d). Given gross key findings and corresponding WSIs, a forensic pathologist can propose a set of potential outcomes as textual queries for any unseen subject. SongCi then predicts the final diagnostic results, providing detailed explanatory factors that underscore the critical elements associated with these predictions. SongCi's effectiveness is validated across a spectrum of forensic pathology tasks, including patch-level post-mortem image generation, self-supervised WSI-level segmentation, extensive forensic diagnosis, and cross-modal explainability analysis. Our results show that SongCi surpasses state-of-the-art multi-modal AI models in internal and external cohorts. Moreover, comparative analysis with forensic pathologists at varying expertise levels reveals that SongCi's insights are on par with those of seasoned experts and significantly surpass those of less experienced pathologists. To our knowledge, SongCi represents the first VLM tailored for forensic pathological analysis and the inaugural large-vocabulary CPath model that operates directly on gigapixel WSIs within the forensic science domain.	
	
	\section*{Results}

	\begin{figure}[htbp]
		\centering
		\includegraphics[width=0.9\linewidth]{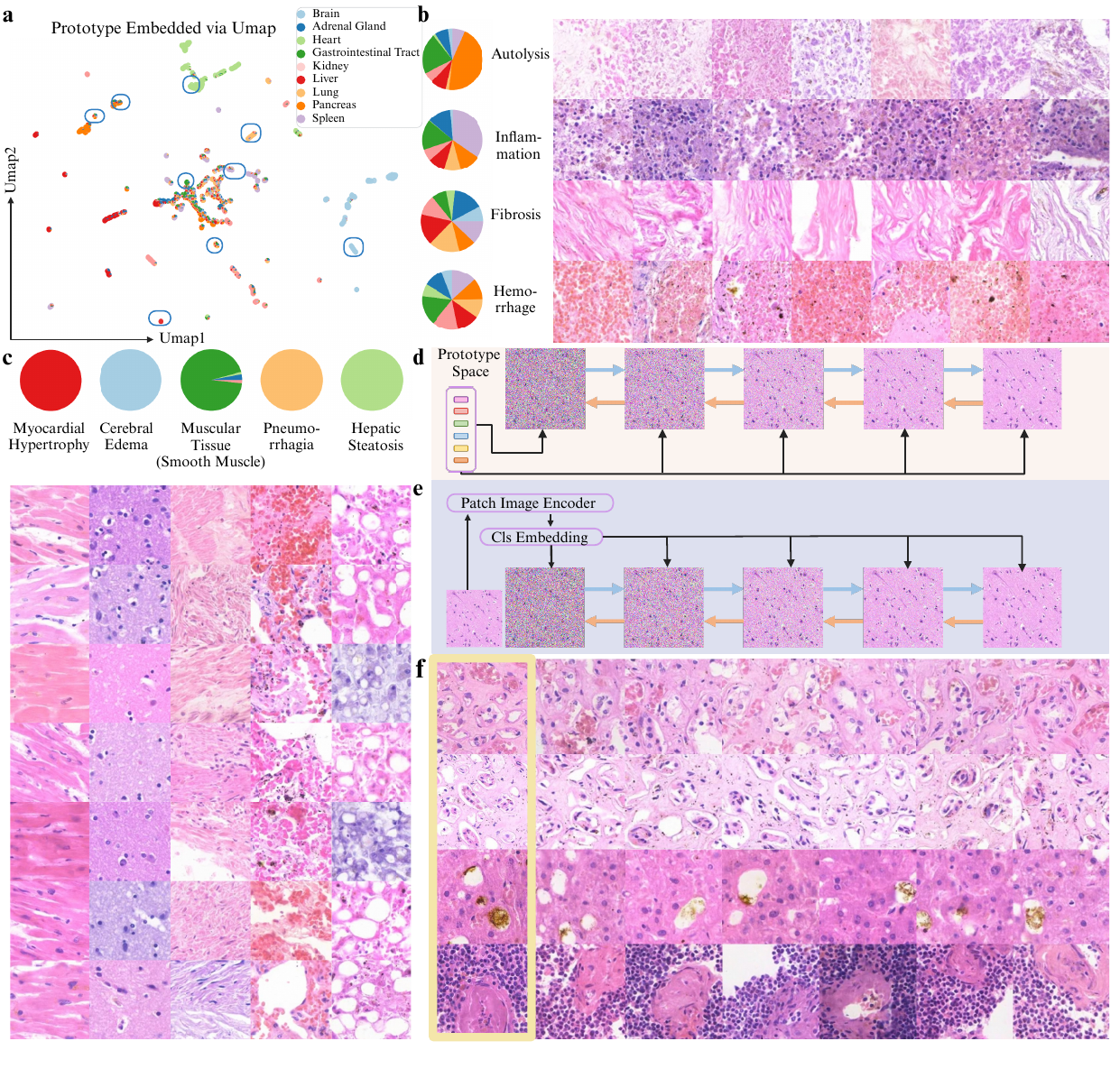}
		\captionsetup{font={tiny,bf,stretch=1.25},justification=raggedright}
		\caption{\textcolor{black}{Prototype representation space of SongCi \& results of post-mortem image generation.
		a. The prototype representation space is visualized using a 2D UMAP, where $933$ dots represent the prototypes, with each dot colored according to the proportion of tissue types it represents.
		b\&c. Prototype-conditioned patch-level generation results. Sub-figure b shows the results for inter-tissue-specific prototypes, including autolysis, inflammation, fibrosis, and hemorrhage. Sub-figure c displays intra-tissue-specific prototypes, such as myocardial hypertrophy, cerebral edema, muscular tissue, pneumorrhagia, and hepatic steatosis.
		d\&e.  The conditional diffusion models are exhibited. Sub-figure d illustrates the prototype-based conditional diffusion model, and sub-figure e shows the instance-based model.
		f. The results of instance-based patch-level generation are presented, featuring representative instances like renal tubules with hemorrhage, normal renal tubules, liver fat particles (undissolved), and splenic trabeculae.}}
		\label{fig:Prototype}
	\end{figure}

\subsection*{Visualization of post-mortem WSI prototypes across different organs}
In a task-agnostic fashion, SongCi employs a prototypical contrastive learning strategy to derive generalizable image representations from post-mortem WSIs of various organs, as depicted in Fig.~\ref{fig:songci}c. Each WSI is segmented into a collection of patches, and an image encoder extracts patch-level representations. These are then projected into a low-dimensional space defined by shared prototypes across WSIs. In our study, we learned a total of 933 prototypes using this SSL method. The organization of these prototypes was visualized using the two-dimensional UMAP technique\cite{umap} and bar plot, as illustrated in Fig.~\ref{fig:Prototype}a and Supplementary Fig. 1, where each dot signifies a prototype, color-coded according to the organ type of the nearest patches. The post-mortem WSIs encompass nine distinct organ types, each represented by a unique color: brain, adrenal gland, heart, gastrointestinal tract, kidney, liver, lung, pancreas, and spleen. Patches from the WSIs are associated with their closest prototype, imparting the corresponding color to the prototype. Figure~\ref{fig:Prototype}a reveals distinct clustering patterns among the prototypes, with some exhibiting uniform colors, denoting intra-tissue prototypes that encode tissue-specific features, such as myocardial hypertrophy and pneumorrhagia (see Fig. ~\ref{fig:Prototype}c). Conversely, prototypes with mixed colors represent inter-tissue prototypes that encapsulate standard histopathological features across different organs, including autolysis, inflammation, fibrosis, and hemorrhage (refer to Fig.~\ref{fig:Prototype}b). These findings suggest that the prototype representations encapsulate both tissue-specific and cross-tissue-shared characteristics from high-resolution WSIs, establishing a versatile foundation for downstream tasks.

\subsection*{Patch-level generation of synthetic post-mortem WSIs}

To assess the generalizability of the prototype and instance representations generated by prototypical contrastive learning, we devised a downstream task focused on patch-level post-mortem image synthesis. We employed representation-conditional diffusion models \cite{rcdm} to generate WSI patches from random noise. These patches were conditioned on the prototypes (see Fig.~\ref{fig:Prototype}d) and instance representations (see Fig.~\ref{fig:Prototype}e). Figure~\ref{fig:Prototype}b displays synthesized images based on four distinct inter-tissue prototypes, which exhibit high-fidelity representations of post-mortem phenomena. The images correspond to various states: extensive autolysis with cellular structure loss, inflammation with abundant lymphocytes, fibrosis, and hemorrhage with dense erythrocytes. These prototypes, learned by SongCi, effectively encapsulate patterns common across different tissues. Additionally, Fig.~\ref{fig:Prototype}c illustrates that SongCi can also capture organ-specific patterns, such as myocardial hypertrophy and cerebral edema, through intra-tissue prototypes. The diffusion model then generates high-fidelity patches for these specific tissue types. The examples in Fig.~\ref{fig:Prototype}b and ~\ref{fig:Prototype}c demonstrate that SongCi's prototypes encode essential post-mortem features of diverse intra-tissues and inter-tissues (more samples see Supplementary Fig. 2). Moreover, Fig.~\ref{fig:Prototype}f showcases image patches generated using specific instance representations as the conditions. These images retain intricate details of the original instances, highlighted by yellow boxes in Fig.~\ref{fig:Prototype}f, including renal tubules with hemorrhage, normal renal tubules, liver fat particles (undissolved), and splenic trabeculae. This indicates that SongCi's instance embeddings are highly detailed, laying a robust groundwork for distilling generalizable prototypes across various organs.

	\begin{figure}[h]
		\centering
		\captionsetup{font={tiny,bf,stretch=1.25},justification=raggedright}
		\includegraphics[width=0.9\linewidth]{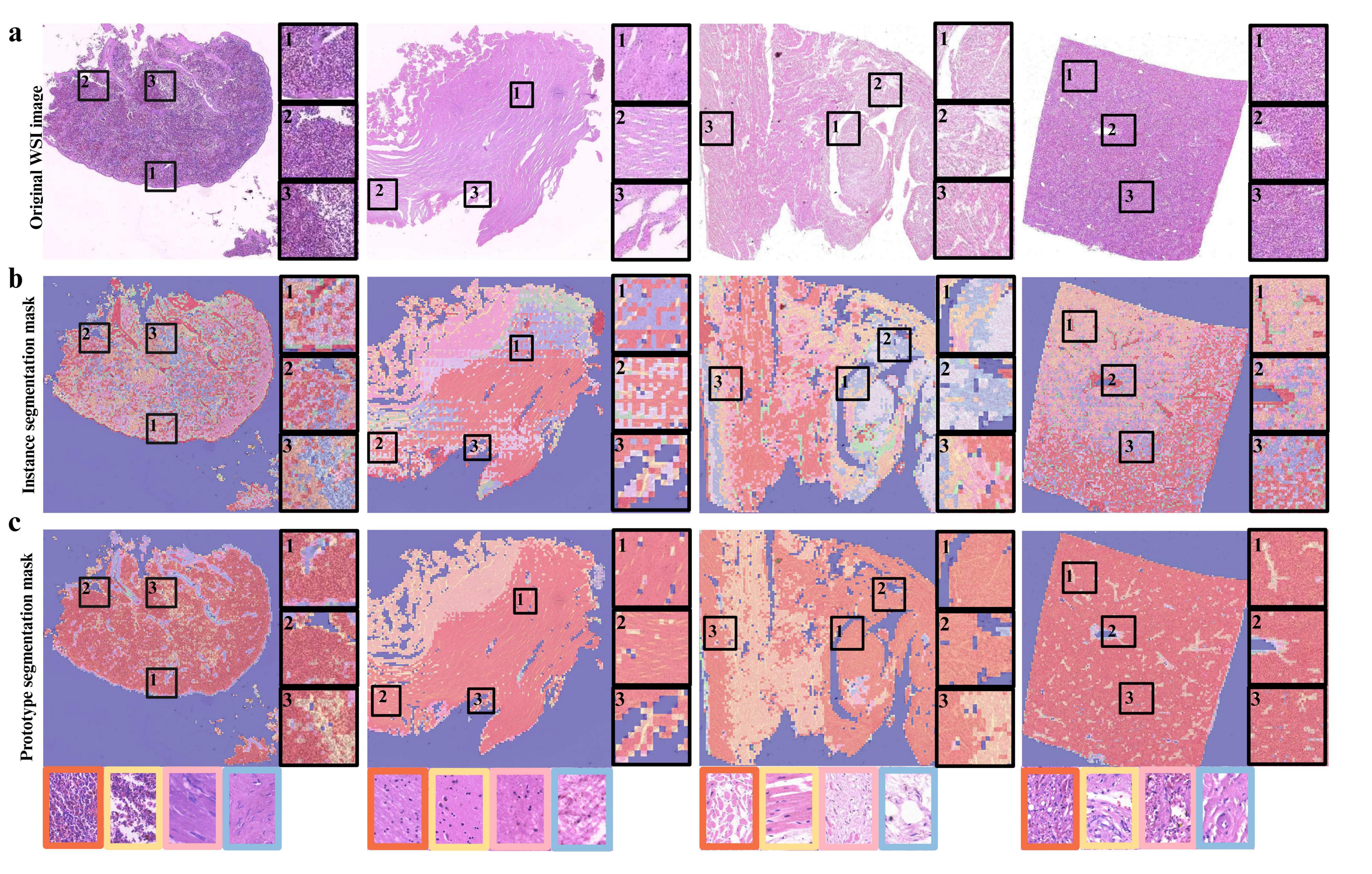}
		\caption{\textcolor{black}{Instance \& prototype segmentation. a displays the original WSIs of four different tissues, including spleen, brain, myocardium, and liver tissues. b and c illustrate the segmentation outcomes utilizing traditional clustering and prototype-based methods, respectively. Each image is partitioned into seven distinct masks, each represented by a unique color based on the quantity of patches within: orange, yellow, pink, blue, white, green, and crimson. The top four prevalent mask types for each image are presented, along with the key distinctions between the two segmentation approaches, which are highlighted with black borders.}}
		\label{fig:seg}
	\end{figure}

\subsection*{Self-supervised segmentation of post-mortem WSIs}

Tissue and cell segmentation are fundamental steps in CPath. Beyond image generation, we have utilized pre-trained prototypical WSI encoders for self-supervised WSI segmentation without fine-tuning. Specifically, we treat each prototype learned by the SongCi as a semantic mask, enabling efficient labeling of all patches in a WSI by matching them with their closest prototypes based on the cosine similarity of their representations. Figure~\ref{fig:seg} showcases self-supervised segmentations of WSIs from four different organs: the spleen, brain, heart, and liver. These are compared with segmentations obtained using an iterative clustering algorithm typically employed in histopathological analysis~\cite{Neuropathologist-level,h2t,Prototypical multiple instance learning}. We can observe that, compared with the conventional iterative clustering (Fig.~\ref{fig:seg}b), SongCi (Fig.~\ref{fig:seg}c) led to much better segmentations in all cases (more samples see Supplementary Fig. 3). The segmentations by SongCi exhibit marked superiority, as evidenced by the precise differentiation between parenchyma and mesenchyme within splenic and hepatic tissues, which are delineated by the red and various other colored sections, respectively. Furthermore, SongCi achieves a level of granularity in brain and myocardial segmentation that aligns closely with cellular structures, minimizing the introduction of noise. In contrast, conventional clustering often results in noisy segmentations that incorrectly separate identical tissues, as indicated by the black boxes in Fig.~\ref{fig:seg}b.


It is worth noting that SongCi offers two significant technical advantages. First, it is more efficient, eliminating the need for iterative optimization and enabling one-shot labeling of patches in ultra-high-resolution WSIs, which may contain over ten thousand patches. Second, it is more flexible and robust, as it does not require pre-determination of cluster numbers. Traditional methods rely heavily on this hyper-parameter, which must be carefully adjusted outside the optimization process. In contrast, SongCi dynamically learns prototypes in a data-driven manner through contrastive SSL. These benefits position SongCi as a generalizable tool for self-supervised or zero-shot segmentation of post-mortem WSIs.

	\begin{figure}[htbp]
		\centering
		\captionsetup{font={tiny,bf,stretch=1.25},justification=raggedright}
		\includegraphics[width=0.9\linewidth]{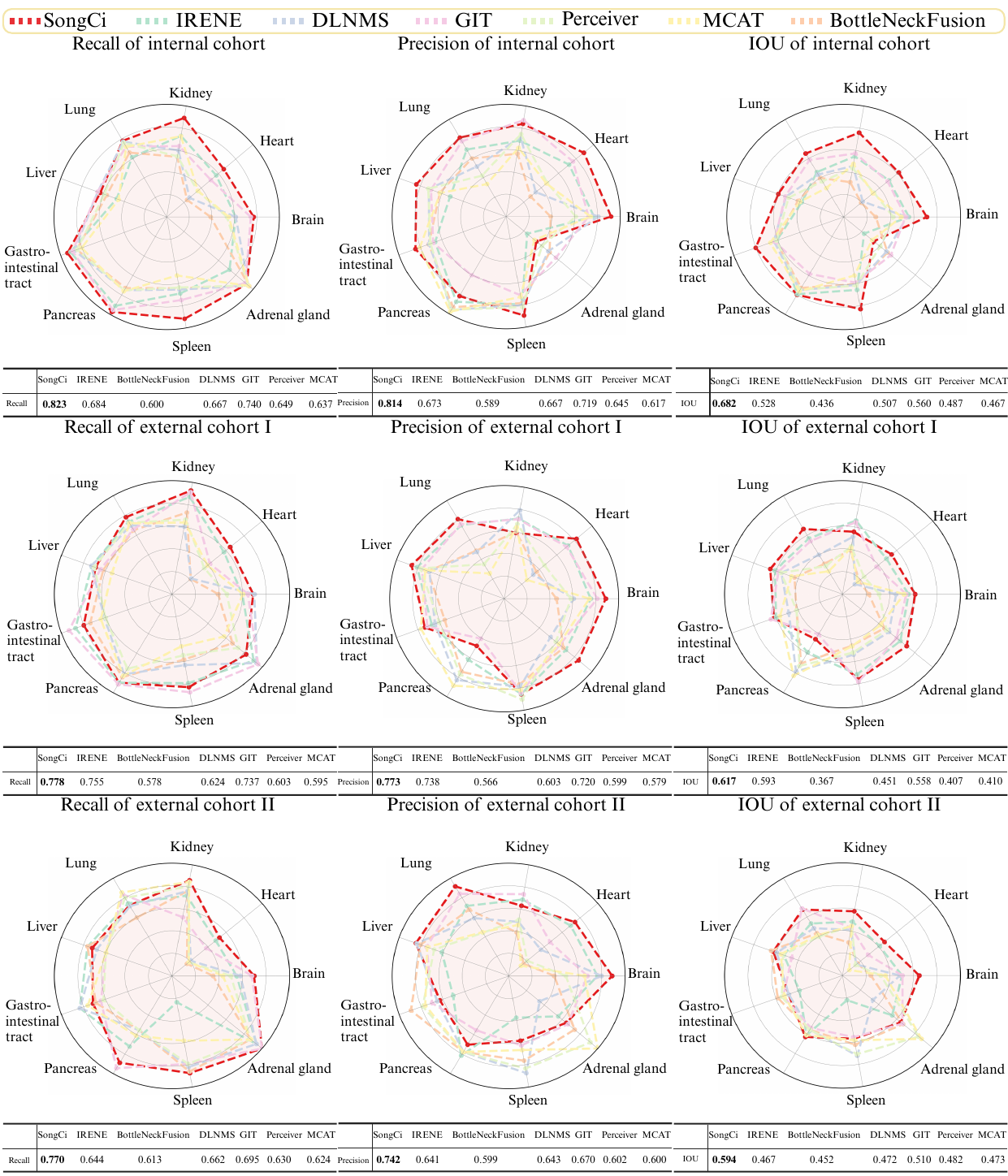}
		\caption{\textcolor{black}{Comparisons of SongCi with state-of-the-art, open-sourced multimodality fusion methods. The evaluation benchmarks SongCi against six established models utilizing three key performance metrics: recall, precision, and Intersection over Union (IoU). Radar charts illustrate the algorithm's efficacy across nine different organs, and the associated table below consolidates the average scores for these organs, with the highest values emphasized in bold.}}
		\label{fig:Comparisons}
	\end{figure}
	\begin{table}[ht]
		\centering
		\fontsize{10}{12}\selectfont
		\begin{threeparttable}
			\caption{\label{tab:tab1}\textbf{Recall of off-set \& low-frequency diagnoses by different methods.}}
				\begin{tabular}{c|cc||ccc}
					\hline
					\multirow{2}{*}{Methods} & \multicolumn{2}{c||}{Off-Set Diagnoses}                     & \multicolumn{3}{c}{Low-Frequency Diagnoses}                   \\ \cline{2-6} 
					& External Cohort I & External Cohort II & Internal Cohort & External Cohort I & External Cohort II \\ \hline 
					Perceiver                &   0.332          &     0.125                             &     0.289        &    0.413         &  0.172           \\
					BottleNeck-Fusion        &   0.300          &           0.146                       &   0.278          &   0.349          &   0.172          \\
					DLNMS                    &   0.259           &          0.167                        &    0.278         &   0.340          &   0.194          \\
					GIT                      &    0.218         &           0.167                       &     0.263        &    0.271         &   0.301          \\
					MCAT                     &    0.318         &           0.250                      &     0.227        &     0.352        &  0.183           \\
					IRENE                    &    0.318         &   \textbf{0.292}                       & 0.344   &     0.413        &  0.333           \\
					SongCi                   & \textbf{0.418}   & \textbf{0.292}                        &    \textbf{0.355}         & \textbf{0.434}   & \textbf{0.366}   \\ \hline
				\end{tabular}
			\begin{tablenotes}
			\fontsize{8}{12}\selectfont
			\item Off-set diagnoses refer to diagnostic events unique to the corresponding cohorts. In contrast, low-frequency diagnoses are diagnostic occurrences that appear fewer than ten times within the cohorts.
			\end{tablenotes}
	\end{threeparttable}
	\end{table}
\subsection*{Large-vocabulary forensic diagnosis, with comparisons to existing VLMs}

Leveraging the pre-trained WSI encoder and a pathology-dedicated language model (i.e., PLIP~\cite{plip}), SongCi designs a cross-modal contrastive learning strategy to learn from multi-modal inputs (i.e., gross key findings and WSIs) a gated-attention-empowered VLM for zero-shot forensic diagnosis, with the schematic diagram shown in Fig.~\ref{fig:songci}e. First of all, we applied this model to the internal cohort and two external cohorts with significantly different data distributions (Fig.~\ref{fig:songci}d), and compared the performance with six state-of-the-art VLMs from both the medical and general machine-learning communities, including IRENE~\cite{irene}, BottleNeckFusion~\cite{bottle-fusion}, DLNMS~\cite{dlnms}, GIT~\cite{git}, Perceiver~\cite{perceiver}, and MCAT~\cite{mcat}. The quantitative results in terms of three metrics (i.e., Recall, Precision, and IOU) obtained by these methods across different organs are summarized in Fig.~\ref{fig:Comparisons} and Supplementary Table. 1. We can see that, on average, SongCi consistently outperformed all other competing VLMs by large margins in terms of all three metrics. Specifically, on the internal cohort, external cohort I, and external cohort II, SongCi got a mean Recall of $0.823\pm0.119$, $0.778\pm0.091$, and $0.770\pm0.132$ , a mean Precision of $0.814\pm0.164$, $0.773\pm0.131$, and $0.742\pm0.117$, and a mean IOU of $0.682\pm0.145$, $0.617\pm0.086$, and $0.594\pm0.070$, respectively. Compared with other VLMs, the average improvements range between $10\%$ and $20\%$ in most cases. To go into more detail, we can observe that SongCi performed significantly better than other VLMs on the two external cohorts in handling post-mortem WSIs from most organs, e.g., brain, heart, lung, and liver, which are closely related to issues such as cause of death determination and manner of death identification.


Furthermore, to comprehensively assess the performance in large-vocabulary forensic pathological analysis, we applied SongCi to the challenging tasks of off-set and low-frequency diagnosis (Table~\ref{tab:tab1}). Specifically, the off-set samples are a subset of each external cohort whose ground-truth diagnostic labels do not exist in the internal cohort for pre-training. The low-frequency samples have labels occurring less than ten times in the corresponding cohorts. Table~\ref{tab:tab1} shows that SongCi consistently led to the best Recall in the off-set and low-frequency diagnosis tasks compared to other VLMs. Notably, these results were obtained by the zero-shot inference via the pre-trained multi-modality models, demonstrating the promising generalizability of SongCi.	

\begin{table}[ht]
	\centering
	\fontsize{10}{12}\selectfont
	\begin{threeparttable}
	\caption{\label{tab:tab4}\textbf{Comparison with human experts on a mini competition test dataset.}}
		\begin{tabular}{c|ccc|c|c}
			\hline
		Operators	& Recall         & Precision      & IOU            & Time (hours) & Feeling (0-4 easy to hard) \\ \hline
			Pathologist Assistant (PA) & 0.593          & 0.494          & 0.403          & 5           & 4                         \\
			Junior Pathologist 1 (JP1) & 0.683          & 0.584          & 0.506          & 15          & 4                         \\
			Junior Pathologist 2 (JP2) & 0.620          & 0.651          & 0.531          & 6.5         & 3                         \\
			Senior Pathologist 1 (SP1) & 0.661          & 0.748          & 0.644          & 8.5         & 1                         \\
			Senior Pathologist 2 (SP2) & { 0.797}    & \textbf{0.878} & \textbf{0.776} & 7           & 0                         \\
			SongCi                    & \textbf{0.836} & { 0.832}    & { 0.712}    & 0.37            & w/o                       \\ \hline
		\end{tabular}
		\begin{tablenotes}
		\fontsize{8}{12}\selectfont
		\item Comparative results involving five forensic pathologists of varying experience levels were evaluated using three distinct assessment metrics. The highest values for each metric are highlighted in bold. Additionally, the pathologists recorded the time taken to complete their assessments and provided subjective evaluations of their completion.
		\end{tablenotes}
\end{threeparttable}
\end{table}		

\subsection*{Comparisons between SongCi and forensic pathologists}
\begin{wrapfigure}{r}{8cm}
	\centering
	\captionsetup{font={tiny,bf,stretch=1.25},justification=raggedright}
	\includegraphics[width=0.4\textwidth]{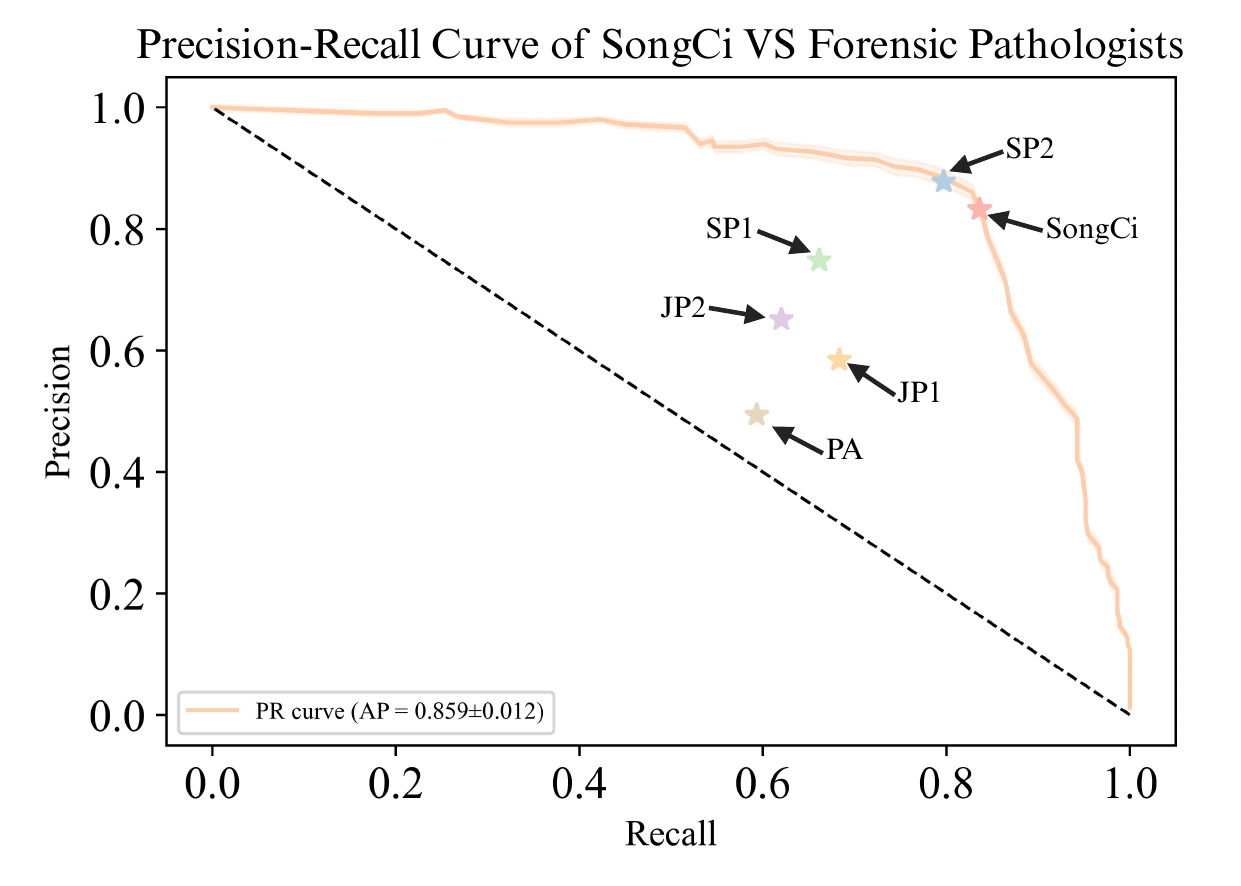}
	\caption[width=0.15\textwidth]{\textcolor{black}{Comparisons of SongCi with human experts. The orange line represents the model's Precision-Recall (PR) curve. Data points on the graph indicate the model's optimal performance alongside the scores of individual pathologists.}}
	\label{fig:human}
\end{wrapfigure}


We compared SongCi with five forensic pathologists with varying expertise levels, i.e., two senior forensic pathologists (SP) with more than 15 years of experience, two junior pathologists (JP) with more than five years of experience, and a pathologist assistant (PA).  Specifically, considering the time-consuming and demanding process of forensic pathology analysis, we selected 100 samples with an unambiguous diagnosis from the two external cohorts and assigned them to these forensic pathologists. We also distributed the internal cohort and ground-truth labels to these experts. Each forensic pathologist analyzed these external samples and made their predictions independently using the internal cohorts as the reference. The results quantified on such an external subset are summarized in Table~\ref{tab:tab4} and Fig.~\ref{fig:human}. From the precision-recall (PR) curve shown in Fig.~\ref{fig:human} and the metric values in Table~\ref{tab:tab4}, we can see that SongCi's performance aligns closely with an SP and significantly surpasses the other SP, two JPs, and the PA. Notably, besides the accuracy matching up with the seasoned pathologists, SongCi outperforms in efficiency (i.e., 0.37 versus 7 hours in Table~\ref{tab:tab4}), which could significantly reduce the workload of forensic pathology.

	\begin{figure}[htbp]
		\centering
		\captionsetup{font={tiny,bf,stretch=1.25},justification=raggedright}
		\includegraphics[width=\linewidth]{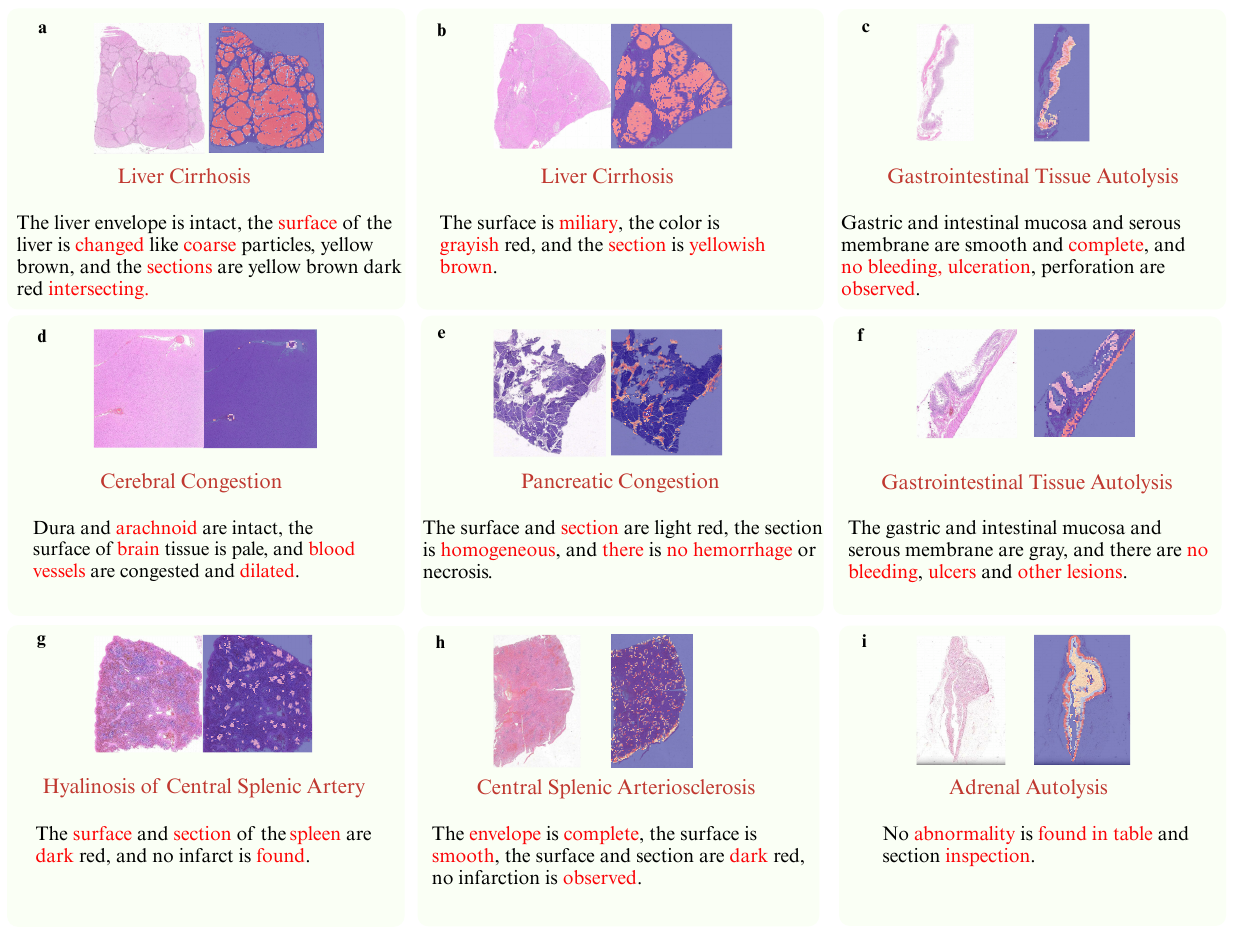}
		\caption{\textcolor{black}{Multi-modality attention visualization of SongCi. The multi-modality attention visualization of SongCi offers interpretable analyses for forensic pathology diagnosis across a range of tissues and organs. Panels a and b display liver tissues; panels c and f, gastrointestinal tissues; panel d, brain tissues; panel e, pancreatic tissues; and panels g and h, spleen tissues, with panel i highlighting adrenal tissues. The WSI regions corresponding to the prototypes of the top five findings, along with the top five vital descriptors in the gross key findings, are delineated in distinct colors.}}
		\label{fig:Interpretation}
	\end{figure}

\subsection*{Multi-modal explainability in forensic pathological analysis} 

SongCi aggregates WSI prototypes and gross key findings to align them with the diagnostic outcomes by learning word/prototype-level attention scores via cross-modal contrastive learning. By nature, these scores also provide critical multi-modal explanation factors for fine-grained analysis of the model's final predictions. In Fig.~\ref{fig:Interpretation} and Supplementary Fig. 4, we show some representative examples across different organs, where, for each case, the top-five prototypes (projected back to the WSIs) and top-five words (located in the text of gross key findings) are highlighted by distinct colors. From Fig.~\ref{fig:Interpretation}, we can have some exciting observations. For instance, as shown in Fig.~\ref{fig:Interpretation}a, the multi-modal attention visualizations regarding liver cirrhosis highlighted meaningful WSI regions and respective words from the gross key findings at the organ level strongly associated with such a specific disease. The WSI's emphasized segment illustrates the typical alterations related to cirrhosis, i.e., pseudolobules. From the gross key findings aspect, the highlighted term `coarse' implies that cirrhosis-affected livers exhibit unevenly sized nodules on their surface, resulting from pseudolobule development, which seems coarse to the unaided eye.  The term ‘intersection’ refers to the overlapping of various colors present on the cirrhosis liver's surface. The terms ‘surface’, ‘section’, and ‘changed’ provide important cues regarding the disease's appearance and position. We can also find that such explanations by SongCi are relatively stable and reproducible, e.g., according to similar results presented in Fig.~\ref{fig:Interpretation}b for the same disease, although the two WSIs look different. Such fine-grained explanations, encompassing textual and visual information, can also be consistently applied to other samples. For instance, regarding the gastrointestinal tissue depicted in Figure~\ref{fig:Interpretation}c, WSI emphasizes areas with autolysis, predominantly in the glandular regions, which exhibit varying degrees of brightness corresponding to the extent of autolysis. In gross anatomical descriptions, the term 'complete' denotes the intactness of the gastrointestinal mucosa and serosa. Conversely, the absence of typical lesions during gross examination is indicated by terms such as 'no', 'bleeding', 'ulceration', and 'observed'. These terminologies reflect the standard normal condition of the gastrointestinal tract in gross anatomy, characterized by the presence of autolytic pathological features alone. These results suggest the generalizability of SongCi from the multi-modal explainability perspective. It demonstrates that the attention-mapping operator of SongCi could be a reliable tool to assist pathologists in post-mortem analysis, helping enhance their confidence and assessment outcomes in forensic investigation.

	\begin{table}[htbp]
		\centering
		\fontsize{10}{12}\selectfont
		\begin{threeparttable}
		\caption{\label{tab:ablation1}\textbf{Ablation study of SongCi with different text encoders.}}

					\begin{tabular}{c|cccccccccccc}
						\hline
						\multirow{2}{*}{Text Encoder} & \multicolumn{4}{c}{Internal Cohort}                                         & \multicolumn{4}{c}{External Cohort I}                                       & \multicolumn{4}{c}{External Cohort II}                              \\ \cline{2-13} 
						& AP             & recall         & precison       & IOU & AP            & recall         & precison       & IOU & AP            & recall         & precison       & IOU            \\ \hline 
						PubmedBERT        &0.364            &0.475           &0.449           &0.315                     &0.326          &0.424          &0.411           &0.239                     &0.405          &0.501           &0.475           &0.323          \\
						CLIP             &0.730            &0.748           &0.728           &0.572                     &0.624          &0.657          &0.655           &0.499                     &0.663          &0.675           &0.670           &0.500         \\
						BiomedCLIP           &0.685            &0.701           &0.682           &0.500             &0.624          &0.656          &0.635          &0.478                     &0.658          &0.653           &0.646           &0.484       \\
						Quilt-1M       &{0.799}            &{0.782}           &{0.780}           &{0.632}                    &{0.749}          &{0.743}          &{0.728}          &{0.572}                     &\textbf{0.763}          &\textbf{0.775}           &{0.738}           &\textbf{0.608}    \\
						PLIP            &\textbf{0.864}            &\textbf{0.823}           &\textbf{0.814}           &\textbf{0.682}          &\textbf{0.808}          &\textbf{0.778}          &\textbf{0.773}           &\textbf{0.617}                     & {0.749}       &{0.770}           & \textbf{0.742}          &{0.594}           \\ \hline
					\end{tabular}

		\begin{tablenotes}
			\fontsize{8}{12}\selectfont
			\item The results of the comparison of various language models under the 4 evaluation metrics in the 3 cohorts, with the highest value in each evaluation metric highlighted.
		\end{tablenotes}
		\end{threeparttable}
		\end{table}
		
		\begin{table}[htbp]
			\centering
			\fontsize{11}{13}\selectfont
			\begin{threeparttable}
				\caption{\label{tab:ablation2}\textbf{Ablation study of WSI prototype encoder with different regularizations.}}

						\begin{tabular}{l|cccccccc}
							\hline
							\multirow{2}{*}{Regularization}                    & \multicolumn{4}{c}{External Cohort I}                                                      & \multicolumn{4}{c}{External Cohort II}                                                      \\ \cline{2-9} 
							& AP                   & Recall               & Precison             & IOU                  & AP                   & Recall               & Precison             & IOU                  \\ \hline
							$L_{ins}+L_{pro}$                            & 0.751                & 0.743                & 0.736                & 0.578                & 0.643                & 0.687                & 0.676                & 0.504                \\
							$L_{ins}+L_{pro}+L_{ipc}+L_{ipd}$            & 0.765 & 0.760 & 0.743 & 0.585 &0.683  &0.695  &0.680  &0.524  \\
							$L_{ins}+L_{pro}+L_{me-max}$                 & 0.773                & 0.765                & 0.750                & 0.602                & 0.691                & 0.701               & 0.684                & 0.527                \\
							$L_{ins}+L_{pro}+L_{me\text{-}max}+L_{ipc}+L_{ipd}$ & \textbf{0.808}                & \textbf{0.778}                & \textbf{0.773}                & \textbf{0.617}                & \textbf{0.749}                & \textbf{0.770}                & \textbf{0.742}                & \textbf{0.594}                \\ \hline
						\end{tabular}
		

			\begin{tablenotes}
				\fontsize{8}{12}\selectfont
				\item The results, derived from the baseline model, i.e.  $L_{ins}+L_{pro}$, incrementally include the addition of regularization terms $L_{ipc}$, $L_{ipd}$, and $L_{me\text{-}max}$.
		\end{tablenotes}
		\end{threeparttable}
		\end{table}

\subsection*{Ablation studies of SongCi's vision and language encoders}

SongCi has two critical components: a pre-trained text encoder to encode gross key findings and a vision encoder trained via prototypical contrastive learning for WSI embedding. The selection of the text encoder and the training strategy of the vision encoder determines the quality of textual and imaging representations, thus the effectiveness of multi-modal fusion. We conducted ablation experiments regarding our specific implementations of SongCi. 


\textit{First}, in SongCi, PLIP~\cite{plip} was used as the text encoder, a language model pre-trained on paired pathology image-text collected from Twitter. We replaced it with the following alternatives: PubmedBERT~\cite{pubmedbert}, i.e., an encoder pre-trained on medical text data only; CLIP~\cite{clip}, i.e., an encoder pre-trained on natural image-text pairs; BiomedCLIP~\cite{biomedclip}, i.e., an encoder pre-trained on medical paper illustrations for image-text pairing, and QUILT-1M~\cite{quilt}, an encoder pre-trained on the pathology image-text pairs grasped from YouTube. The large-vocabulary forensic diagnoses obtained by these variants are summarized in Table~\ref{tab:ablation1}, from which we can have two observations. That is, we can see that foundation models pre-trained by image-text pairs (e.g., CLIP, BiomedCLIP, QUILT-1M, and PLIP) perform better than those that use only texts (e.g., PubmedBERT), implying the merit of multi-modal learning. In addition, we can see that the two pathology-dedicated LLMs (i.e., QUILT-1M and PLIP) outperformed others by large margins, demonstrating the power of domain knowledge in cross-modal information alignment and fusion. Such domain knowledge encoded in texts is critical in improving generalizability and reducing the reliance on large training datasets. The PLIP used in SongCi led to the best performance in most cases.
	

\textit{Second}, to pre-train the vision encoder by task-agnostic SSL, SongCi designed sophisticated regularization terms to constrain the prototypical space. To check the efficacy of these regularization terms, we further conducted an ablation study to remove them from the loss function and quantify the influence on the diagnosis performance, with the results summarized in Table~\ref{tab:ablation2}. As can be seen, the baseline (i.e., the model pre-trained by using $L_{ins}+L_{pro}$) has the worst performance, without the three regularization terms on the prototypical space (i.e., $L_{me\mathrm{-}max}$, $L_{ipc}$, and $L_{ipd}$). The inclusion of these terms significantly improved the diagnostic outcomes. This demonstrates the effectiveness of the constraints designed in SongCi for learning complete prototypical space, allowing the encoder to establish better the mapping relationship between instance embeddings and prototypical embeddings.	
	
\section*{Discussion}

This study presents a generalizable and explainable vision-language model (VLM), i.e., SongCi, dedicated to forensic pathology. To build SongCi, we curated one of the largest post-mortem, multi-modal datasets, gathering 2,228 paired WSI-text samples from three forensic centers, nine organs, and 471 diagnostic outcomes. By leveraging cutting-edge self-supervised learning techniques, the pre-trained SongCi was evaluated on a broad spectrum of downstream tasks in forensic pathological analyses, demonstrating exciting accuracy, generalizability, and explainability compared with state-of-the-art VLMs and forensic pathologists. Our research addresses a significant gap in the availability of multimodal AI tools for complex diagnostic and predictive tasks in forensic pathology. This discipline has traditionally depended on expert judgment and is marked by subjectivity, inconsistency, and inefficiency. 

A primary challenge in building forensic VLMs is extracting and aligning multimodal representations from challenging post-mortem data for large-vocabulary analyses. For this purpose, one major technical strength of SongCi lies in customizing a prototypical contrastive learning algorithm to pre-train a powerful image encoder for fine-grained feature extraction from post-mortem WSI with atypical and varying appearances. In a task-agnostic fashion, it maps millions of patches from WSIs into a low-dimensional space spanned by limited prototypes, where similar instances or patches are grouped tightly from intrinsic semantic views across different organs. The joint visualization of the learned prototypical and instance representations via the UMAP technique provides an intuitive way to understand their organization and relationships among various tissue types. Such visualization revealed that SongCi learns to partition the complex post-mortem WSI space into interpretable clusters corresponding to distinct histopathological entities or disease states, either inter-tissue shared or intra-tissue specific. Two downstream tasks, post-mortem image generation at the patch level and self-supervised semantic segmentation at the WSI level, evaluated the utility of the pre-trained image encoder. The image generation results demonstrate that SongCi's pre-trained image encoder can provide strong guidance to boost advanced diffusion models, generating controllable image patches with high fidelity that present detailed post-mortem changes and organ-specific lesions. Based on the learned prototypes, the self-supervised semantic segmentation results show that a gigapixel WSI can be efficiently segmented as meaningful and essential areas of interest under remarkable precision. These evaluations suggest the generalizability and reliability of SongCi's image encoder as a CPath tool in forensic pathology, even without fine-tuning.

To seamlessly align multi-modal, post-mortem data, another key innovation of SongCi is the design of a cross-modal contrastive learning algorithm to establish a dedicated fusion block empowered by gated-attention mechanisms. It plays a pivotal role in integrating macroscopic observations at the organ level (i.e., gross key findings) with microscopic cues at the tissue level (i.e., WSIs) to produce higher-level representations encoding multi-modal knowledge for accurate and coherent forensic pathological analyses. The downstream applications demonstrate that this design led to much better large-vocabulary diagnostic performance than other open-sourced VLMs in forensic pathology. More importantly, its accuracy has matched that of experienced forensic pathologists while significantly improving efficiency, further implying the meaning of SongCi in the autopsy practice. The gated-attention mechanisms in this fusion block assign relative importance or scores to each input element, i.e., each patch in a WSI and each word in the textual description of the gross key findings, by which SongCi straightforwardly focuses on the most salient aspects of both modalities. This mimics the workflow of a forensic pathologist, who typically makes his/her judgments by analyzing organ-level autopsy findings in conjunction with microscopic assessments. The results of explainability analysis demonstrate that such attention mechanisms can capture fine-grained cross-modal factors to uncover how SongCi makes a particular prediction given specific multi-modal inputs, which is very critical considering that AI tools for forensic pathology by nature have exceptionally high requirements in reliability and trustiness.

Strong zero-shot learning capability is a standout advantage of SongCi, indicating its adaptability and generalizability to novel scenarios, a featured challenge in forensic pathology, considering that autopsies typically involve a broad array of investigations across multiple organs and varying conditions. The success of this zero-shot transfer learning capability lies in SongCi's robust cross-modal fusion and alignment layer, which effectively combines information across modalities to create an aligned representation space for both textual and visual data. Specifically, in the inference stage, given the post-mortem WSI and gross key findings of a particular subject, an operator can list a set of suspicious diagnoses (in texts) as candidates, which could be new cases that have not yet been seen in the pre-training stage. Then, SongCi calculates the cosine similarity between the multi-modal fusion representations and the embeddings of the provided candidate diagnoses, based on which the most likely outcome can be ranked out, together with detailed explanation factors pinpointing specific aspects of the multi-modal data that significantly influence the model's decision. This empowers a forensic pathologist to assess in detail the relationships between a given sample and various potential diagnoses by providing a quantitative measure of confidence, thus assisting pathologists in making accurate assessments and potentially reducing diagnostic errors or inconsistencies. The large-vocabulary forensic diagnostic results across the two external cohorts, especially the off-set and low-frequency quantifications presented in Table~\ref{tab:tab1}, demonstrate the superior zero-shot learning performance of SongCi. Furthermore, the comprehensive comparisons with the existing VLMs and forensic pathologists show that SongCi is a generalizable, explainable, and, more importantly, forensic pathology-dedicated AI tool, adept at efficiently amalgamating various data sources, thereby enhancing the efficiency, accuracy, and consistency of forensic diagnoses across varying cases and organs.

This work has several limitations that deserve continual research in the future. Although an unprecedentedly large dataset of paired post-mortem WSIs and gross key findings was collected for the self-supervised learning of SongCi, more data with significant diversities are needed to improve its robustness and generalizability further. Considering that data collection in forensic pathology is practically more complicated than in clinical applications, and the former has more obvious large-vocabulary properties, national or even international collaborations on this topic are urgently necessary. The current version of SongCi is a VLM, which could neglect critical forensic information in other data modalities. In practice, an autopsy includes multiple steps, which produce a broad spectrum of data formats like textual, multi-omics, and imaging data. Intuitively, the fusion of this multi-modal information could further improve the outcomes in AI-empowered forensic pathology. Although we have justified the efficacy of SongCi through a series of downstream tasks, some other applications still need further investigation, such as predicting post-mortem time and simulating longitudinal post-mortem changes conditioned on varying organs, environments, and death causes. A more comprehensive design of downstream evaluations is vital in enhancing the practical usage of SongCi. Moreover, using English-based LLMs may limit SongCi's applicability in non-English communities, necessitating adaptations for broader adoption.


In summary, both the presented applications and SongCi's existing limitations pronounce the significance of continuous research and evaluation to advance and better understand the strengths and practical usage of cross-modal self-supervised pre-training (or even so-called multi-modal foundation models) for forensic pathology.

\section*{Methods}

\subsection*{SongCi}

SongCi is a multi-modal deep learning model tailored for forensic pathological analyses. The architecture consists of three main parts: an imaging encoder for WSI feature extraction, a text encoder for embedding gross key findings and diagnostic queries, and a multi-modal fusion block that integrates the embeddings of WSI and gross key findings to align with those of the diagnostic queries. Specifically, we used an open-sourced, pathology-dedicated language model, i.e., PLIP, in SongCi as the textual encoder directly. To deal with post-mortem data with varying conditions, we designed two novel self-supervised learning (SSL) algorithms to build the imaging encoder and multi-modal fusion block in a task-agnostic fashion. In inference, SongCi can flexibly conduct large-vocabulary (or even open-vocabulary) diagnosis, as an operator only needs to provide a set of candidate outcomes, based on which the model ranks out the possible diagnosis associated with detailed explanation factors identified from the multi-modal inputs.

\subsubsection*{Prototypical WSI encoder} 

We propose a hybrid contrastive learning algorithm to learn from gigapixel, post-mortem WSIs fine-grained representations generalizable across different organs (Fig.~\ref{fig:songci}c). The algorithm is built upon a straightforward assumption that image patches (i.e., instances) from different spatial locations, organs, and conditions, are grouped as meaningful clusters in the desired representation space that captures both intra-tissue-specific and inter-tissue-specific information; besides, in each cluster, instance representations present a certain degree of variance to preserve detailed patch-wise specificity. Accordingly, such a hybrid SSL algorithm consists of an instance contrastive learning part and a prototypical contrastive learning part.

The instance contrastive learning part aims to build a vision transformer (ViT)~\cite{vit}, which leverages the local patches of a WSI as the input and learns instance-level representations via self-distillation. 
In line with DINO~\cite{dino}, contrastive learning is achieved by a teacher-student strategy. 
A batch of (say totally $N$) instances is first transformed by a series of data augmentation operations, including multi-cropping~\cite{swav}, random resizing, flipping, color jittering, solarization, and Gaussian blurring, which produces two different views for each input instance (say $X_s^n$ and $X_t^n$ for the n-th instance). 
Then, $X_s^n$ and $X_t^n$ are fed into the student and teacher branches, respectively, and we want the corresponding predictions by the two branches to be cross-view consistent. 
Specifically, the student branch consists of a ViT $f_s(\cdot)$, a projector $g_s^{pro}(\cdot)$, and a predictor $g_s^{pre}(\cdot)$. 
Compared with the student branch, the teacher branch also contains a ViT $f_t(\cdot)$ and a projector $g_t^{pro}(\cdot)$, while the last component is replaced by a sharpening \& centering module $J_t(\cdot)$, which adjusts the distribution of the instance representations to avoid mode collapse. 
The model parameters of the student branch are optimized by gradient back-propagation, based on which the teacher branch is updated via the exponential moving average (EMA) \cite{moco}.  
To this end, we quantify the predictions from the two branches via soft-max normalization, such as equation (1) and equation (2)

$$p_s(X_s^n) = \frac{exp(g_s^{pre}(g_s^{pro}(f_s(X_s^n))) / \tau_s)}{\sum_{n=1}^N exp(g_s^{pre}(g_s^{pro}(f_s(X_s^n))) / \tau_s)} (1)$$
$$p_t(X_t^n) = \frac{exp(J_t(g_t^{pro}(f_t(X_t^n))) / \tau_t)}{\sum_{n=1}^N exp(J_t(g_t^{pro}(f_t(X_t^n))) / \tau_t)} (2)$$

where $\tau_s$ and $\tau_t$ are two hyperparameters that control the sharpness of the output probability distributions, respectively. 
After that, the model parameters are iteratively optimized by minimizing the cross-entropy loss that encourages cross-view consistency, such as equation (3)

$$L_{ins} = -\sum_n p_t(X_t^n)log(p_s(X_s^n)) (3)$$

The prototypical contrastive learning part further distills from the instance representations a more abstract and generalizable feature space spanned by a set of learnable prototypes shared across different WSIs, organs, and conditions. 
Similar to the instance contrastive learning, here we have two different views for an input instance after data augmentation; the difference in this prototypical learning procedure is that the learnable parameters of the student branch are updated via back-propagation and the teacher branch shares weights with the student branch. 
Both branches contain a ViT followed by a projector. Let their output embeddings for N input instances be $Z_s \in \mathbb{R}^{D\times N}$ and $Z_t \in \mathbb{R}^{D\times N}$, respectively. 
We want to learn a set of M prototypical embeddings, say $P \in \mathbb{R}^{D\times M}$, for which we can find a linear mapping $C_t \in \mathbb{R}^{M*N}_+$ that maximizes their similarities with $Z_t$, i.e., minimizes the Sinkhorn distances of the associated optimal-transport problem~\cite{Sinkhorn}, defined as equation (4) and equation (5)

	$$\mathop{\max}_{C_t}Tr(C_t^TP^TZ_t) + \epsilon h(C_t) (4)$$
	
\begin{center}
	\textit{s.t.}
	$ C_t\in \mathbb{R}^{M*N}_+, C_t^T1^M=1^N,C_t1^N=\frac{N}{M}1^M (5)$
\end{center}

where $1^N$ and $1^M$ denote the N- and M- dimensional all-ones vectors, respectively. 
The function $Tr(\cdot)$ stands for the matrix trace operation, $h(C_t) = -\sum_{ij} C_t[i,j]log(C_t[i,j])$ quantifies the entropy of the linear mapping, and $\epsilon$ is a tuning parameter controls its influence.
Given $P$ and $Z_t$, a $C_t$ qualified for the above objective realizes that all instances in $Z_t$ can be matched up to the prototype space P and each prototype is selected at least $\frac{N}{M}$ times on average.
Such a constrained optimization problem can have an approximate solution by using the iterative Sinkhorn-Knopp algorithm~\cite{swav,Sinkhorn}, with each iteration defined as equation (6):
 
$$\hat{C}_t=diag(u)exp(\frac{P^TZ_t}{\epsilon})diag(v) (6)$$

where $u\in \mathbb{R}^{M}$ and $v\in \mathbb{R}^{N}$ are re-normalization vectors, and $diag(\cdot)$ formulate them as the diagonal matrices. 
Following SwAV~\cite{swav}, the number of iterations was set as three in our study.

To establish the cross-view consistency, we use $\hat{C}_t$ from the teacher branch as the pseudo label to constrain the prediction in the student branch. 
That is,  the student branch predicts the mapping matrix as equation (7)

$$C_s = \frac{exp(P^TZ_s / \tau_s^{prototype})}{\sum_{m=1}^M exp(P^TZ_s / \tau_s^{prototype})} (7) $$

where $ \tau_s^{prototype}$ is a element-wise scaling parameter. Then, the main loss function for the prototypical contrastive learning is defined as equation (8)

	$$L_{prototype} = - \sum \hat{C}_t log(C_s) (8)$$
	
Furthermore, to stabilize the learning of the prototypical representations, we design three regularization terms in addition to the main loss function.	
Ideally, the prototype-instance mapping matrix should be sparse rather than dense, by which the prototypes are encouraged to encode diverse information and each instance tends to have the most similar prototype(s).
To this end, we impose an instance-prototype cross-entropy loss ($L_{ipc}$) and an instance-prototype distance loss ($L_{ipd}$), which are defined as equation (9), equation (10) and equation (11).

	$$L_{ipc}= -\sum_{i=1}^N log\left(\frac{exp(z_i^Tp_j^i)}{\sum_{j=1}^M exp(z_i^Tp_j)}\right) (9)$$
	
	$$L_{ipd} = \sum_{i=1}^N(z_i - p_j^i )^2 = \sum_{i=1}^N (2-2z_i^Tp_j^c) (10)$$
	
	$$p_j^i = \mathop{max}\limits_{p_j}(\{z_i^Tp_j\}_{j=1}^M) (11)$$
	
where $z_i$ and $p_j$ denote the $L_2$-normalized $D$-dimensional representation of the $i$-th instance and $j$-th prototype, respectively, and $p_j^i$ is the nearest prototype of $z_i$.
As a result, $L_{ipc}$ encourages inter-prototype differences, and $L_{ipd}$ encourages intra-cluster consistency, i.e., instances close to a particular prototype should be grouped tightly. 
In addition, to obtain fine-grained representations by the prototypes, we encourage the prototype-instance mapping to fully use all these prototypes, for which a mean-entropy maximization regularization~\cite{msn} is attached, such as equation (12)

	$$L_{me\text{-}max} = \sum _{i=1}^M\bar{C_s^i}log(\bar{C_s^i}) + log(M) (12)$$
	
where $\bar{C_s^i} = \frac{1}{N} \sum _{j=1}^N C_s[i,j]$. Therefore, the global loss for the hybrid contrastive learning is defined as equation (13)

	$$L_{final} = \lambda_1 L_{ins} + \lambda_2 L_{prototype} + \lambda_3 L_{ipc} + \lambda_4 L_{ipd} + \lambda_5 L_{me\text{-}max} (13)$$
	
where $\lambda_1$ was set as $0.6$, $\lambda_3$ was $0.1$, and $\lambda_2=\lambda_4=\lambda_5=1.0$ in our implementation. 
After the contrastive learning, the ViT, the projector, and the prototypes are frozen to be the pre-trained prototypical WSI encoder. All the patches' nearest prototypes determine the WSI-level representation.

\subsubsection*{Gated-attention-boosted multi-modal fusion block} 

We propose a multi-modal fusion block empowered by gated attention mechanisms for the adaptive fusion of gross key findings and WSI information, which produces multi-modal representations encoding macroscopic and microscopic cues for large-vocabulary forensic pathological analyses.

This gated-attention-boosted fusion block adopts initial imaging embeddings of a WSI and textual embeddings of the paired gross key findings as the multi-modal input. 
The pre-trained prototypical WSI encoder produces the initial imaging embeddings. 
That is, each patch of a WSI is denoted by the nearest prototype according to the cosine similarity between instance and prototypical representations, and the combination of all patches' nearest prototypes forms the WSI-level prototypical \textit{feature embedding}, say $z^i_{pro}$. 
Moreover, it is worth noting that the frequency of a particular prototype occurring in a WSI could encode critical information regarding the WSI's specific patterns. 
For example, in a pathology section of brain autolysis, the proportion of normal brain tissue varies due to the degree of autolysis, which can be reflected by the frequency of the post-mortem autolysis-related prototype selected by the pathological image. 
To encode such key information, we further quantify the numbers of occurrences of each prototype for a WSI and use them as the WSI's prototypical \textit{num embedding}, say $z^i_{num}$. 
On the other hand, the initial word-wise textual embeddings of gross key findings (say $z^t$) are produced by PLIP, a pathology-dedicated LLM. 
To better align PLIP with forensic pathology, we attach a simple but effective adaptation layer, i.e., a text-to-image adapter $f_{adapter}(\cdot)$ onto PLIP, which fine-tunes $z^t$ in a few-shot knowledge transfer fashion~\cite{clip-adapter,tip-adapter}. 
Furthermore, considering that the initial mono-modal feature embedding brings inevitable information loss that could cause a cross-modal mismatch, we follow C-MCR~\cite{cmc} to update these initial embeddings by adding a small amount of Gaussian noise to improve their robustness for subsequent cross-modal fusion. 
More specifically, given the initial embeddings of the paired WSI (i.e., $z^i_{pro}$ and $z^i_{num}$) and gross key findings (i.e., $z^t$), they are first refined before feeding into the fusion block, such as equation (14) and equation (15)

$$\hat{z}^i =\frac{z^i_{pro}+\epsilon^i}{\Vert z^i_{pro}+\epsilon^i \Vert_2} + z^i_{num} (14)$$

$$\hat{z}^t =f_{adapter}(\frac{z^t+\epsilon^t}{\Vert z^t+\epsilon^t \Vert_2}) (15)$$ 

where $\epsilon^i$ and $\epsilon^t$ denote the random Gaussian noises, and $\Vert\cdot\Vert_2$ stands for $L_2$-normalization.

Given the refined multi-modal inputs (i.e., $\hat{z}^i$ and $\hat{z}^t$), the fusion block designs gated cross-attention and feed-forward network (FFN) layers to update the representations of each modality by considering the complementary information from the other modality.  
The gated mechanism~\cite{flamingo} adaptively controls the inter-modal information transfer, thus balancing the contributions of different modalities during cross-modal communication, which has been proven to have typically better performance than conventional cross-attention strategies~\cite{mcat,irene}. 
Specifically, by using $\hat{z}^t$ as the guidance, the WSI embedding $\hat{z}^i$ is updated by the gated knowledge-guided cross-attention layer followed by a gated FFN, which can be formulated as equation (16) and equation (17)

$$\hat{z}^i_l = \hat{z}^i + tanh(\lambda^i_{att}) \cdot softmax(\frac{W_q \hat{z}^i(W_k \hat{z}^t)^T}{\sqrt{d_k}})W_v\hat{z}^t (16)$$

$$\hat{z}^i_{l+1} =\hat{z}^i_l +tanh(\lambda^i_{ffn}) \cdot FFN_i(\hat{z}^i_l) (17)$$

where $W_q$, $W_k$ , and $W_v$ stand for learnable matrices for linear mappings, $\lambda^i_{att}$ and $\lambda^i_{ffn}$ are learnable scalars (i.e., the gated-attention coefficients), and $tanh(\cdot)$ denotes the tanh activation. 
Similarly, by using $\hat{z}^i$ as the guidance, $\hat{z}^t$ is updated by the gated prototype-guided cross-attention layer followed by a gated FFN; equation (18) and equation (19):

	$$\hat{z}^t_l = \hat{z}^t + tanh(\lambda^t_{att}) \cdot softmax(\frac{W_q \hat{z}^t(W_k \hat{z}^i)^T}{\sqrt{d_k}})W_v\hat{z}^i (18)$$
	
$$\hat{z}^t_{l+1} =\hat{z}^t_l +tanh(\lambda^t_{ffn}) \cdot FFN_t(\hat{z}^t_l) (19)$$

where $\lambda^i_{att}$ and $\lambda^i_{ffn}$ are the respective gated-attention coefficients.

After that, we concatenate $\hat{z}^t_{l+1}$ and $\hat{z}^i_{l+1}$, and apply a gated Transformer-based encoder (with two layers) to update these representations. 
They are further processed by a modality projector to obtain the multi-modal representations, say $z^f\in \mathbb{R}^{D*M}$, where $M$ denotes the number of tokens, and $D$ stands for feature dimensionality.

Considering the large-vocabulary property of forensic pathological analysis, the flexibility in aligning the multi-modal representation $z^f$ with the diagnostic outcomes is a key issue.
Practically, an autopsy investigation could lead to multi-label outcomes from different views (or according to different downstream needs). 
For instance, given a post-mortem liver case, one could have the diagnoses of liver autolysis, cirrhosis, and rupture from the perspectives of post-mortem changes, diseases, and injury patterns, respectively.
Thus, the multi-modal representation $z^f$ should be able to seamlessly align with each of them without bias to any particular outcome, a common limitation of many existing VLMs, like CLIP~\cite{clip,open-seg,Perceptual Grouping,cdul}. 
It is intuitive to assume that different diagnostic outcomes (after the embedding by a text encoder) are associated with varying parts of $z^f$, based on which we design an adaptive alignment strategy to flexibly align between the representations of multi-modal inputs and large-vocabulary outcomes.
Specifically, any particular diagnostic outcome is regarded as a caption of the corresponding case and is mapped by the frozen PLIP and a learnable linear projector to obtain its caption embedding, say $z^d \in \mathbb{R}^{D*1}$. 
By calculating the cosine similarities between the normalized $z^d$ and $z^f$ across different tokens, we can obtain the attention scores (say $z^s \in \mathbb{R}^{1*M}$) indicating the contribution of each token to this diagnosis, such as equation (20) :
   
$$ z^s = \frac{exp((\frac{z^d}{\Vert z^d \Vert_2})^T \frac{z^f}{\Vert z^f \Vert_2})}{\sum_{i=1}^M exp((\frac{z^d}{\Vert z^d \Vert_2})^T \frac{z^f_i}{\Vert z^f_i \Vert_2})} (20)$$

where $\Vert\cdot\Vert_2$ stands for $L_2$-normalization.
Finally, we leverage $z^s$ to aggregate $z^f$, yielding $\hat{z}^f = z^f (z^s)^T$ to align with $z^d$.
Notably, changing $z^d$ leads to changing $z^s$, thus different $\hat{z}^f$.

Finally, given $\hat{z}^f$ and $z^d$ for training samples, we minimize the \textit{InfoNCE loss}~\cite{infonce} to update the learnable parts of this multi-modal fusion block, such as equation (21), equation (22) and equation (23):

	$$L_{f \rightarrow d} = - \sum_{i=1}^{N} log \frac{exp((\frac{\hat{z^f_i}}{\Vert \hat{z^f_i} \Vert_2})^T \frac{z^d_i}{\Vert z^d_i \Vert_2}/ \lambda)}{\sum_{j=1}^{N}exp((\frac{\hat{z^f_i}}{\Vert \hat{z^f_i} \Vert_2})^T \frac{z^d_j}{\Vert z^d_j \Vert_2}/ \lambda)} (21)$$
	
	$$L_{d \rightarrow f} = - \sum_{i=1}^{N} log \frac{exp((\frac{\hat{z^f_i}}{\Vert \hat{z^f_i} \Vert_2})^T \frac{z^d_i}{\Vert z^d_i \Vert_2}/ \lambda)}{\sum_{j=1}^{N}exp((\frac{\hat{z^f_j}}{\Vert \hat{z^f_j} \Vert_2})^T \frac{z^d_i}{\Vert z^d_i \Vert_2}/ \lambda)} (22)$$
	
	$$L_{all} = \frac{L_{f \rightarrow d} + L_{d \rightarrow f}}{2} (23)$$
	
where $N$ is the batch size and $\lambda$ is the scaling temperature parameter.

\subsubsection*{Inference process} 


The pre-trained SongCi model serves as an auxiliary tool for forensic pathologists, facilitating diagnosis and analysis through zero-shot learning~\cite{clip}. When presented with a case that includes multi-modal inputs, such as WSIs and macroscopic gross findings from various organs, a forensic pathologist can propose multiple diagnostic hypotheses from different perspectives, including post-mortem alterations, disease classifications, and injury patterns. SongCi then computes organ-specific multi-modal feature embeddings and evaluates their cosine similarity with each proposed diagnosis. The model ranks the most probable diagnoses based on normalized similarity scores, applying a predefined threshold to identify the top outcomes. In our research, we established the threshold at 0.88, corresponding to the point on the precision-recall (PR) curve where precision and recall are optimally balanced. In addition to the possible diagnostic results, fine-grained interpretable analytical factors are also available to forensic pathologists. This process enables them to scrutinize the interpretable results and refine their assessments.

\subsubsection*{Implementation details}


A total of 3.15 million patches were extracted from the WSIs(the internal cohort) to train the prototypical whole slide image encoder. The network's backbone was initialized using DEiT-small~\cite{deit} and subsequently fine-tuned over 250 epochs with a mini-batch size of 768. Both the instance and prototype projectors ($projector^{ins} \& projector^{pro}$) and the instance predictor ($predictor^{ins}$) share an identical architecture, comprising two fully connected layers, a batch normalization layer\cite{bn}, and a Gaussian error linear unit (GELU) layer\cite{gelu}. These components were trained from scratch using Xavier initialization. The AdamW\cite{adamw} optimization algorithm was employed, starting with an initial learning rate of $1e-6$ and concluding with a final rate of $1e-7$. A cosine annealing strategy was applied to decay the learning rate. The prototype vectors were initially set to a dimensionality of 1,024, with their parameters being frozen at the outset and later included in the training after 5,000 iterations. Post-training, each patch of WSIs in the internal cohort was matched to the most similar prototype in the prototype space using only resizing for data augmentation. Upon reviewing all 3.15 million patches, it was found that 933 prototypes had corresponding patches.


The multi-modal fusion layer underwent training for 500 epochs, utilizing a mini-batch size of 64. The architecture comprises one text-to-image adapter, two cross-attention layers, two feedforward neural networks (FFNs), one transformer encoder, and one modality projector layer. The depth of the transformer encoder is 2. Following the settings in CLIP, the word embedding length is set to 77, while the prototype embedding length is 128. Embeddings are truncated or padded to maintain these specified lengths. To improve the generalization and robustness of the fusion block, random noise drawn from a normal distribution with a mean of 0 and a standard deviation of 0.5 was introduced to the embeddings (word and prototype). All training and experimental procedures were performed on a PC Server with eight NVIDIA GEFORCE RTX 3090 GPUs.

\subsection*{Prototype visualization \& patch-level post-mortem WSI generation}	

We adopted the Umap method~\cite{umap} to visualize in 2-D the prototypes learned by SongCi, as shown in Fig.~\ref{fig:Prototype}a. 
Each prototype was marked by one or multiple color(s) according to the ratio of organ types of nearest image patches. To check the generalizability of the patch and prototypical embeddings produced by SongCi, we conducted a downstream patch-level post-mortem image generation task. 
Specifically, a diffusion model~\cite{gdiffusion} was trained to generate image patches conditioned on a prototype or patch embedding. Consistent with DDPM~\cite{ddpm} and DDIM~\cite{ddim}, the training of our diffusion model contains a forward process to add noises and a reverse process that learns to predict added noises. 
In the noise prediction, we incorporate conditions by combining learned embedding and time embedding:

$$tc = f(e_{time-embedding})+g(e_{learned-embedding}) (24)$$

$$ x_t = \sqrt{\alpha_t}x_0 +\sqrt{1-\alpha_t} \epsilon  (25)$$

$$ x_{t-1} = \sqrt{\alpha_{t-1}}(\frac{x_t - \sqrt{1-\alpha_t}\epsilon_\theta(x_t,tc)}{\sqrt{\alpha_t}})+\sqrt{1 - \alpha_{t-1}-\delta_t^2}\epsilon_\theta(x_t,tc)+\delta_t\epsilon (26)$$

where $\epsilon \sim N(0,I)$,$\epsilon_\theta$ is a Unet~\cite{unet},  and $f(\cdot)$ and $g(\cdot)$ are two fully connected networks. 
The hyperparameter $\delta_t$ controls the generation process~\cite{ddim}, and $\alpha_t$ controls the accumulation of noise intensity at the moment ($t$) of the diffusion process.

\subsection*{Self-supervised WSI segmentation \& diagnostic explainability analysis}

	
Based on the prototypes produced by SongCi, we conducted a downstream task of self-supervised post-mortem WSI segmentation. Specifically, we assume that each prototype encodes specific semantic information, and then each patch of a WSI was assigned a semantic label according to its nearest prototype. Notably, for a particular WSI, if the resulting prototype categories exceed a predefined threshold, we further cluster\cite{kmeans} the number of prototypes to be equal to the threshold for comparison with other methods. In our experiments, the threshold was set as seven.


In the zero-shot diagnosis process, for a candidate outcome, the cross-modal fusion block of SongCi assigns respective attention scores to each patch of the input WSI as well as each word of the gross key findings, which provide fine-grained, cross-modal explanations regarding the network prediction. Specifically, we visualized the patches and words with the top five attention scores in our experiments.
	
\subsection*{Comparative analysis study} 
In our study, SongCi was compared with six state-of-the-art multi-modal fusion methods in the medical domain. For a fair comparison, all these methods were implemented under the same configurations with SongCi, such as the batch size and number of epochs, etc. Consistent with SongCi, these multimodal fusion methods adopt the embeddings of a WSI and respective gross key findings as the multimodal input, from which they learn fused representations in different ways to align with the diagnostic outcomes. These competing methods include:
\begin{itemize}
	\item  Multimodal Co-Attention Transformer (MCAT)~\cite{mcat}: MCAT is a transformer-based model that designs a genomic-guided co-attention layer to fuse multimodal information. In this study, we replaced the input of genomic embedding with gross key findings. Other operations remained the same as in the original implementation of MCAT.
	\item Generative Image-to-text Transformer(GIT)~\cite{git}: This model uses a specific text decoder to fuse multimodal embeddings via multiple self-attention layers. For a fair comparison, we trained such a model using the same loss functions of SongCi.	
	\item IRENE~\cite{irene}: As an extension of GIT, IRENE designed a bidirectional multimodal attention block for cross-modal information fusion. We change the image patch tokens to WSI image prototype tokens and clinical text tokens representing key gross findings.
	\item Perceiver~\cite{perceiver}: The input of the Perceiver has two parts: a latent array and a byte array. The byte array stores input multi-modal embeddings and the latent array is a learnable embedding with random initialization. Based on the byte array, the model updates the latent array by alternatively calling self-attention and cross-attention layers. After that, the learned latent array is output as the fused representation.	
	\item Bottleneck-Fusion~\cite{bottle-fusion}: The original Bottleneck-Fusion method was proposed to combine WSI and tabular clinical data to predict lymph node metastasis of papillary thyroid carcinoma~\cite{bottle-shared}. We replaced tabular clinical data with the textual description of gross key findings for forensic pathological analyses.	
	\item DLNMS~\cite{dlnms}: As a classic late-fusion method, DLNMS directly concatenates multi-modal embeddings and feeds them into an attention-based, fully connected network for diagnosis or prediction. The original model was developed for occult nodal metastasis prediction. We kept the methodological design but applied it to our task of forensic pathological diagnosis.		
	
\end{itemize}

	\subsection*{Forensic pathology cohorts}
	This work included three cohorts: an internal cohort provided by the Forensic Judicial Expertise Center of Xi'an Jiaotong University and two external cohorts provided by Shaanxi Zhongjin Judicial Expertise Center and Shanghai Academy of Forensic Science, China. The data collection procedures satisfied the requirements of local laws and were approved and supervised by the Ethics Committee of the corresponding institutions. The internal cohort contains 164 decedents from 2018 to 2023, for which a total of 1,451 paired samples of gross key findings and WSIs (together with corresponding pathological diagnoses) were collected. The external cohort I (Shaanxi Zhongjin Judicial Expertise Center) contains 50 decedents with a total of 467 gross key findings - WSIs - forensic pathological diagnoses. The external cohort II (Shanghai Academy of Forensic Science) contains 14 deceased individuals with a total of 310 gross key findings - WSIs - forensic pathological diagnoses. These data were from nine different organs, including the brain, heart, lung, kidney, liver, pancreas, spleen, adrenal gland, and gastrointestinal tract (see Fig. \ref{fig:songci}a and \ref{fig:songci}b). The gross key findings (text) represent the forensic pathologist's description of the organ's condition at autopsy, encompassing organ-level information. WSIs (image) are sections derived from regions of interest (ROIs) selected by the forensic pathologist from the deceased's organs, capturing tissue-level information. The forensic pathological diagnoses (text) detail specific outcomes for each organ, typically addressing trauma, disease, and post-mortem changes. Thus, each sample contains multiple forensic diagnoses (multi-labels).

	\subsection*{Data preprocessing} 
	

	\subsubsection*{Forensic pathology slides (WSIs)}
	
	In the analysis of pathology slides, the initial step involves the extraction of regions containing tissue. Utilizing the\textit{finding contours} method in \href{https://opencv.org/}{OpenCV}, we delineate the edges of pathological tissue within a WSI. The process begins by converting the WSI's RGB image at 10x magnification to a binary image using a threshold of $200$. Subsequently, we employ the findContours method to generate a hierarchical tree-structured contour. We iterate through all contours, retaining those with an area exceeding 100,000, which we classify as foreground information. This preliminary contour allows us to differentiate between WSI foreground and background, resulting in a mask image. The mask image guides the segmentation patch process, wherein we preserve patches with more than 40\% foreground information. Each patch is archived as a PNG file with dimensions of $256\times256$ pixels, facilitating subsequent analytical tasks. Due to the nature of forensic pathology, which typically features a high tissue density per slide, the number of patches extracted can range from tens to hundreds of thousands per slide.

	\subsubsection*{Gross key findings}

	During an autopsy, the forensic pathologist systematically examines each organ and documents the autopsy's gross key findings in detail. Our study aimed to address two main issues in response to the gross anatomical findings. Firstly, the challenge of language diversity: the existing open-source medical foundational language models, such as PLIP, QUILT, and BioMedClip, only accept English input, necessitating the translation of texts into English. Secondly, the anatomical records often contain text of varying lengths with extra information; hence, we sought to standardize the key findings' text to a uniform size, facilitating the foundational model's learning process. Leveraging ChatGPT's\cite{gpt} capabilities, this preprocessing was achieved using a prompt-based learning approach. The prompt instructed: '\textit{As a professor in the field of artificial intelligence in forensic medical imaging, translate the following paragraph into English, summarize it into no more than four sentences, and ensure it adheres to academic writing standards. Provide only the summarized content, omitting any extraneous information.}' Subsequently, two forensic pathologists reviewed and validated the processed text, with their amendments constituting the final version.

	\subsubsection*{Forensic pathology diagnosis}	
	
	Forensic pathological diagnosis is conducted by meticulously examining an organ's gross anatomy and histopathology. This discipline primarily focuses on three aspects: trauma, disease, and postmortem changes. Consequently, a single organ may have multiple forensic pathological diagnoses, in contrast to oncologic pathology, which primarily determines the presence or absence of a tumor. From a deep learning perspective, forensic pathology diagnosis represents a complex detection task characterized by a broad spectrum of labels and a substantial quantity of labels rather than a mere classification task. In our study, we engaged three forensic pathologists to consolidate the pathological diagnoses according to the aforementioned tripartite framework. These diagnoses were then compiled into phrases and depicted using a word cloud (see Fig. \ref{fig:songci}B). For instance, a forensic pathological diagnosis of brain tissue might include cerebral congestion (indicative of postmortem changes), cerebral contusion (suggestive of trauma), and brain degeneration (signifying disease).

	\subsection*{Data availability} 

	The internal cohort in this study is publicly available merely for academic research. Out of concern for the deceased's privacy, we only provide content closely related to this study. The rest of the information is not available to the public now.
	
	\subsection*{Code availability} 
	
	\textcolor{black}{The training and inference scripts, and trained models have been publicly released. Please refer to \href{https://github.com/shenxiaochenn/SongCi}{https://github.com/shenxiaochenn/SongCi} for more details, including the explanations of forensic pathology diagnoses and gross key findings.	}

\section*{Acknowledgements}

The authors acknowledge the funding of the Council of the National Natural Science Foundation of China (No. NSFC81730056).

\section*{Author contributions}
C.S. and C.L. conceived the study and wrote the manuscripts.
C.S. and W.Z. developed the SongCi model and wrote the code.
C.S., F.W., and K.L. performed the experiment analysis.
X.W., G.W., H.W., and X.L. preprocessed the datasets.
S.F., J.Z., and H.M. provided the datasets.
C.L., J.M., and Z.W. supervised the project.

\section*{Competing interests}

The authors declare no competing interests.

\section*{Additional information}

\subsection*{Supplementary information}

For additional information, please refer to the supplementary-information.docx

\section*{Figure Legends}

\begin{itemize}
	\item  \textbf{Figure 1} \textbf{The framework of SongCi and studied large-vocabulary, multi-center datasets.}
					
	\textbf{a} Overview of WSI data. ~The dataset spans a broad spectrum of samples from nine various human organs, each meticulously annotated.
	\textbf{b} Data Structure and Provenance. ~The dataset was compiled from three premier forensic cohorts. Diagnostic outcomes in forensic pathology were represented using a word cloud and a Venn diagram to illustrate the distribution and overlap of diagnoses.
	\textbf{c} Training process of WSI prototype encoder. ~SongCi utilizes a self-supervised contrastive learning framework, augmented with prototype-based clustering strategies to enhance efficiency, forming the basis of the prototypical WSI encoder.
	\textbf{d} Inference process of SongCi. ~SongCi processes the gross anatomical key findings and WSIs to generate a potential diagnosis. Additionally, SongCi provides a diagnostic rationale by highlighting significant terms related to the gross key findings and identifying suspicious regions within the WSIs.
	\textbf{e} Training process of cross-multimodality fusion and alignment layer. ~Within the SongCi framework, diverse data modalities, including gross key findings and WSIs, are integrated using an innovative gated-attention-boosted multimodal fusion block. Subsequently, the framework aligns the unified representation space with forensic pathological diagnoses through self-supervised contrastive learning, effectively establishing inter-modal correlations.
	For more detailed information, please refer to the Methods section.
	\item \textbf{Figure 2} \textbf{ Prototype representation space of SongCi \& results of post-mortem image generation.}
	\textbf{a} The prototype representation space is visualized using a 2D UMAP, where $933$ dots represent the prototypes, with each dot colored according to the proportion of tissue types it represents.
	\textbf{b\&c} Prototype-conditioned patch-level generation results. Sub-figure b shows the results for inter-tissue-specific prototypes, including autolysis, inflammation, fibrosis, and hemorrhage. Sub-figure c displays intra-tissue-specific prototypes, such as myocardial hypertrophy, cerebral edema, muscular tissue, pneumorrhagia, and hepatic steatosis.
	\textbf{d\&e}  The conditional diffusion models are exhibited. Sub-figure d illustrates the prototype-based conditional diffusion model, and sub-figure e shows the instance-based model.
	\textbf{f} The results of instance-based patch-level generation are presented, featuring representative instances like renal tubules with hemorrhage, normal renal tubules, liver fat particles (undissolved), and splenic trabeculae.
	\item \textbf{Figure 3} \textbf{Instance \& prototype segmentation.} 
	\textbf{a} displays the original WSIs of four different tissues, including spleen, brain, myocardium, and liver tissues. 
	\textbf{b \& c} illustrate the segmentation outcomes utilizing traditional clustering and prototype-based methods, respectively. Each image is partitioned into seven distinct masks, each represented by a unique color based on the number of patches within orange, yellow, pink, blue, white, green, and crimson. The top four prevalent mask types for each image are presented, along with the key distinctions between the two segmentation approaches, which are highlighted with black borders.
	\item \textbf{Figure 4} \textbf{Comparisons of SongCi with state-of-the-art, open-sourced multimodality fusion methods.}
	The evaluation benchmarks SongCi against six established models utilizing three key performance metrics: recall, precision, and Intersection over Union (IoU). Radar charts illustrate the algorithm's efficacy across nine different organs, and the associated table below consolidates the average scores for these organs, with the highest values emphasized in bold.
	\item \textbf{Figure 5} \textbf{Comparisons of SongCi with human experts.} 
	The orange line represents the model's Precision-Recall (PR) curve. Data points on the graph indicate the model's optimal performance alongside the scores of individual pathologists.
	\item \textbf{Figure 6} \textbf{Multi-modality attention visualization of SongCi. }
	The multi-modality attention visualization of SongCi offers interpretable analyses for forensic pathology diagnosis across a range of tissues and organs. Panels a and b display liver tissues; panels c and f, gastrointestinal tissues; panel d, brain tissues; panel e, pancreatic tissues; and panels g and h, spleen tissues, with panel i highlighting adrenal tissues. The WSI regions corresponding to the prototypes of the top five findings, along with the top five vital descriptors in the gross key findings, are delineated in distinct colors.

\end{itemize}

\end{document}